\pdfoutput=1
\documentclass[11pt]{article}

\usepackage[final]{nlp4MusA}
\usepackage{booktabs}
\usepackage{multirow}
\usepackage{times}
\usepackage{latexsym}
\usepackage{adjustbox}
\usepackage{placeins}

\usepackage[T1]{fontenc}
\usepackage[utf8]{inputenc}

\usepackage{microtype}

\usepackage{inconsolata}
\usepackage{booktabs}
\usepackage{multirow}
\usepackage{colortbl}
\usepackage{graphicx}
\usepackage{booktabs} % For better horizontal rules
\usepackage{caption} % For better table captions
\usepackage{tabularx}
\usepackage{array}
\usepackage{booktabs} % For better horizontal rules
\usepackage{caption} % For better table captions
\usepackage{xcolor} % For coloring the table cells

\title{PIAST: A Multimodal Piano Dataset with Audio, Symbolic and Text}

\usepackage{authblk}

\author[1]{Hayeon Bang}
\author[1]{Eunjin Choi}
\author[1]{Megan Finch}
\author[1]{Seungheon Doh}

\author[2]{\\Seolhee Lee}
\author[2]{Gyeong-Hoon Lee}
\author[1]{Juhan Nam}

\affil[1]{Graduate School of Culture Technology, KAIST, South Korea}
\affil[2]{NCSOFT, South Korea}

\affil[ ]{\small \texttt{\{hayeonbang, \hspace{-0.3em}jech, \hspace{-0.3em}meganelisabethfinch, \hspace{-0.3em}seungheondoh, \hspace{-0.3em}juhan.nam\}@kaist.ac.kr,}\hspace{0.3em}
\texttt{\{seolhee, \hspace{-0.3em}ghlee0304\}@ncsoft.com}}

\begin{document}
\maketitle
\begin{abstract}
While piano music has become a significant area of study in Music Information Retrieval (MIR), there is a notable lack of datasets for piano solo music with text labels. To address this gap, we present PIAST (PIano dataset with Audio, Symbolic, and Text), a piano music dataset. Utilizing a piano-specific taxonomy of semantic tags, we collected 9,673 tracks from YouTube and added human annotations for 2,023 tracks by music experts, resulting in two subsets: PIAST-YT and PIAST-AT. Both include audio, text, tag annotations, and transcribed MIDI utilizing state-of-the-art piano transcription and beat tracking models. Among many possible tasks with the multimodal dataset, we conduct music tagging and retrieval using both audio and MIDI data and report baseline performances to demonstrate its potential as a valuable resource for MIR research.

\end{abstract}

\section{Introduction}

Piano music presents unique opportunities for music research due to its ability to express diverse styles using a single instrument and its superior transcription performance. Given these characteristics, it has become a significant area of study in Music Information Retrieval (MIR), encompassing tasks such as classification \cite{emopia, midibert}, and music generation with various conditions \cite{compose, pop2piano}. While these tasks require datasets that combine piano audio with various modalities such as MIDI, sheet music, or text, there is a notable scarcity of such comprehensive multimodal piano datasets.

However, existing multimodal music datasets, particularly music-text datasets, rarely focus exclusively on piano music, and piano solo pieces comprise only a small portion of general music-text datasets. For instance, in the ECALS Dataset~\cite{toward}, a subset of the Million Song Dataset \cite{msd}, the number of piano solo tracks is very limited. We observed that excluding tracks tagged with instruments other than the piano or genres that could not be solely represented by the piano, only approximately 0.46\% of the entire dataset can be identified as piano solo music.

Several piano datasets, such as MAESTRO \cite{maestro}, have been developed in recent years, which provide classical piano performances primarily used for piano transcription. Another classical piano dataset, GiantMIDI \cite{giantmidi}, is also commonly used in transcription tasks. Other datasets like Pop1K7 \cite{compound} focus on the performance generation of pop piano music, while PiJAMA \cite{pijama} is employed for performer identification tasks with their jazz piano data.
% Transcrtiption-based 말고 POP909와 같이 전문가가 직접 pop song을 피아노로 편곡한 데이터셋도 존재한다.
%There are also datasets that are not transcription-based, such as POP909 \cite{pop909-ismir2020}%, which is a piano arrangement of 909 Chinese pop songs by professionals.
However, these datasets are confined to a single genre and lack text labels. This absence of genre diversity within a single dataset and the lack of textual information underscores the need for a piano dataset with text information.

Some piano datasets contain emotion labels, such as EMOPIA \cite{emopia} and VGMIDI \cite{vgmidi}. However, these datasets are annotated only with emotion information based on either Russell's four quadrants \cite{emopia} or the valence-arousal model \cite{vgmidi}. This limited annotation approach lacks the rich textual descriptions needed for text-based MIR tasks.

To address the limitations, we present multimodal piano music data with rich text annotations and transcribed MIDI. To build the dataset, we first created a piano-specific taxonomy with 31 tags that include genre, emotion, mood, and style information to encompass the broad and diverse musical range that the piano can express. Based on this taxonomy, we collected data from YouTube, transcribed it to MIDI format, and conducted an annotation process.

The PIAST dataset consists of two subsets: \textbf{PIAST-YT}, 9,673 tracks collected from YouTube, providing audio and text information (titles, tags, and descriptions), and \textbf{PIAST-AT}, 2,023 tracks with annotations by music experts. This dual approach ensures both breadth and accuracy in the dataset. Additionally, PIAST includes transcribed performance MIDI data alongside audio and text, enhancing its capabilities beyond existing methods \cite{compound}.

This paper details the dataset collection process and analyzes the data. We present baseline results for piano music annotation and retrieval tasks, utilizing the PIAST-YT and the PIAST-AT datasets across audio and MIDI domains. The PIAST dataset is available in our online repository\footnote{\url{https://hayeonbang.github.io/PIAST_dataset/}}, and the source code for the experiments can be found on GitHub\footnote{\url{https://github.com/Hayeonbang/PIAST}}.

%multimodal learning has become crucial in Music Information Retrieval (MIR), enabling nuanced analysis of musical content. Datasets combining audio with textual descriptions facilitate advanced tasks like text-based music retrieval and generation. However, these datasets often lack comprehensive coverage of specific instruments or genres, limiting their applicability in specialized research areas.

%The piano, with its ability to express diverse styles using a single instrument and superior transcription performance, presents a unique opportunity for advancing music research. However, existing music-text datasets rarely focus exclusively on piano music, and piano solo pieces comprise only a small portion of general music-text datasets.

\section{Dataset}\label{sec:dataset}
\subsection{Taxonomy for Piano Music}

To encompass and precisely define the range of expressions possible in solo piano music, we constructed a comprehensive taxonomy considering genre, emotion, mood, and style tags. We classified genres suitable for solo piano music into four categories: jazz, classical, new-age, and pop piano covers, defining sub-genres within each. The detailed classification of the classical genre was not included in this dataset due to the extensive range and complexity unique to classical music. For emotion and mood taxonomy, we combined vocabularies from four existing music datasets with emotion tags \cite{cal500, emotify, deam, ym2413}), eliminating overlaps. Seven music experts who majored in music composition rated the tags on a 1-5 Likert scale for their suitability in describing solo piano music. We included only words scoring 3.5 or higher and established a taxonomy of 39 tags. After the first annotation process, we removed tags with excessively high co-occurrence or low selection frequencies, resulting in a final specialized taxonomy of 31 words for piano music.

\subsection{The PIAST-YT Dataset}
The PIAST-YT dataset comprises approximately 9,673 tracks (1,006 hours) of audio collected from YouTube, accompanied by text information (title, tags, and descriptions of the video). We employed two collection methods: tag-based and channel-based. The tag-based method used our taxonomy to gather diverse styles of piano music from YouTube. However, the inherent variability in the availability of solo piano content on the platform led to some imbalance in the collected data. To ensure the inclusion of high-quality solo piano content, we also employed a channel-based method, collecting piano performance videos from 23 selected channels known for their piano content. Finally, The PIAST-YT dataset comprises three main components after pre-processing step: audio extracted from videos, text data (titles, tags, and descriptions), and MIDI data generated through transcription.

\begin{figure}[t!]
\centering
\includegraphics[width=0.9\columnwidth]{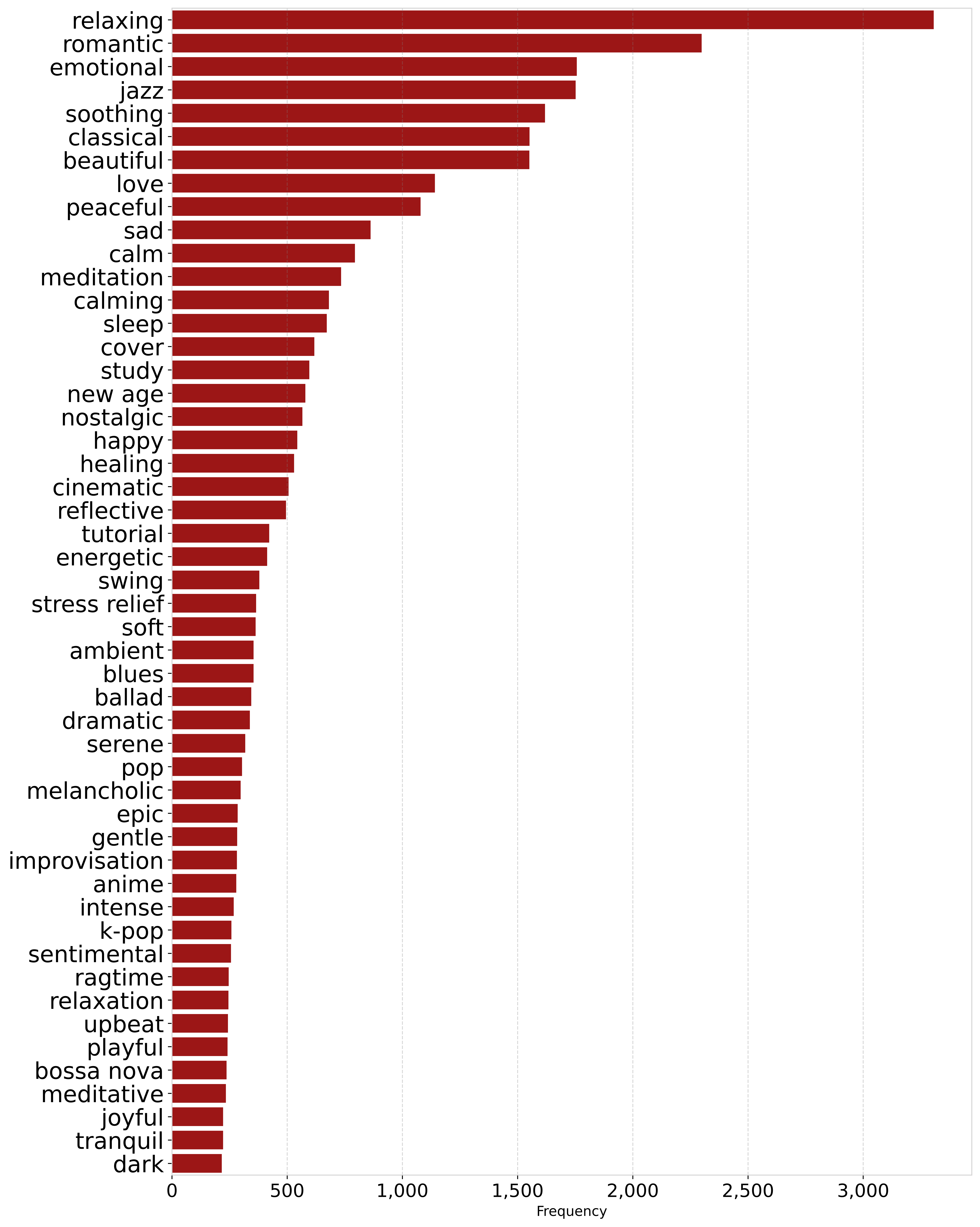}
\vspace{-2.5mm}
\caption{Top 50 words most frequently appearing in the text dataset of the PIAST-YT.}
\vspace{-2.5mm}
\label{fig:yt}
\end{figure}

\subsubsection{Pre-processing}
\textbf{Audio}: To isolate pure piano solo performances, we filtered the data using musicnn \cite{musicnn}, excluding tracks with non-piano sounds in their top 5 tags. Files exceeding 2 hours were removed, and those exceeding 30 minutes were segmented into 10-minute chunks for data consistency. This process reduced the original 1,789 hours of data, about 44\%, to 1,006 hours.\\
\textbf{Text}: The collected text data from YouTube contained diverse and irrelevant information. To extract relevant music-descriptive features, we employed an LLM-based model, specifically ChatGPT 4-Turbo \cite{chatgpt}, chosen for its high performance. This model generated a tag list for each video based on its corresponding text. Figure \ref{fig:yt} illustrates the distribution of these generated tags. The total number of vocabulary is 3,160.\\
\textbf{MIDI}: The piano audio files were transcribed to performance MIDI using an automatic piano transcription model \cite{kong2021high}. The MIDI was then synchronised to beat estimates, and melody and chords were extracted using the Pop1k7 dataset pipeline \cite{compound}. For the beat estimates, following \cite{beatcommittee}, we used the Mean Mutual Agreement (MMA) between a `committee' of several state-of-the-art beat trackers, including All-in-One \cite{kim2023all} and madmom \cite{bock2014madmomDBN1}, to filter out samples for which the beat tracking quality was poor. This transcription process was applied to the audio in both the PIAST-YT and the PIAST-AT datasets.
% \subsubsection{Symbolic Transcription}
%TODO: Symbolic Transcription 
% Beat committee
% Dataset filtering
%The piano audio files were transcribed to performance MIDI using an automatic piano transcription model \cite{kong2021high}. The MIDI was then synchronised to beat estimates, and melody and chords were extracted using the Pop1k7 dataset pipeline \cite{compound}. For the beat estimates, following \cite{beatcommittee}, we used the mean mutual agreement (MMA) between a `committee' of several state-of-the-art beat trackers, including All-in-One \cite{kim2023all} and madmom \cite{bock2014madmomDBN1}, to filter out samples for which the beat tracking quality was poor. This transcription process was applied to the audio in both the PIAST-YT and the PIAST-AT datasets.

\subsection{The PIAST-AT Dataset}
Even after processing, the text data in the PIAST-YT exhibited several limitations. Although it was processed using an LLM-based model, it still showed a low correlation with the music content, and some audio files lacked corresponding text data. To address these issues, we created the PIAST-AT, a dataset consisting of piano-specific human-annotated text.

\subsubsection{Annotation Process}
We stratified 2,400 samples from audio data of PIAST-YT based on the queries used during the collection process, and extracted a 30-second segment from each sample for human annotation. The process involved 15 music experts (7 jazz and 8 classical musicians) with majors in composition. Each segment was assigned to three annotators using a web-based system. The annotators were divided into five groups tagged with 230 segments. Detailed descriptions and examples were provided to all annotators for each tag to ensure consistency. They were also instructed to exclude samples that did not strictly adhere to solo piano criteria or had subtle mood changes. After two rounds of annotation, we collected tags for 2,023 samples (approximately 17 hours of original audio), with 377 samples excluded through this process.
% \begin{figure}
%  \centerline{
%  \includegraphics[width=0.9\columnwidth]{fig/Annotation.png}}
%  \caption{Annotation interface used in the PIAST-AT dataset.}
%  \label{fig:anno}
% \end{figure}

\subsubsection{Dataset Analysis \& Tag Consensus}
\begin{figure}[!t]
 \centerline{
 \includegraphics[width=1.0\columnwidth]{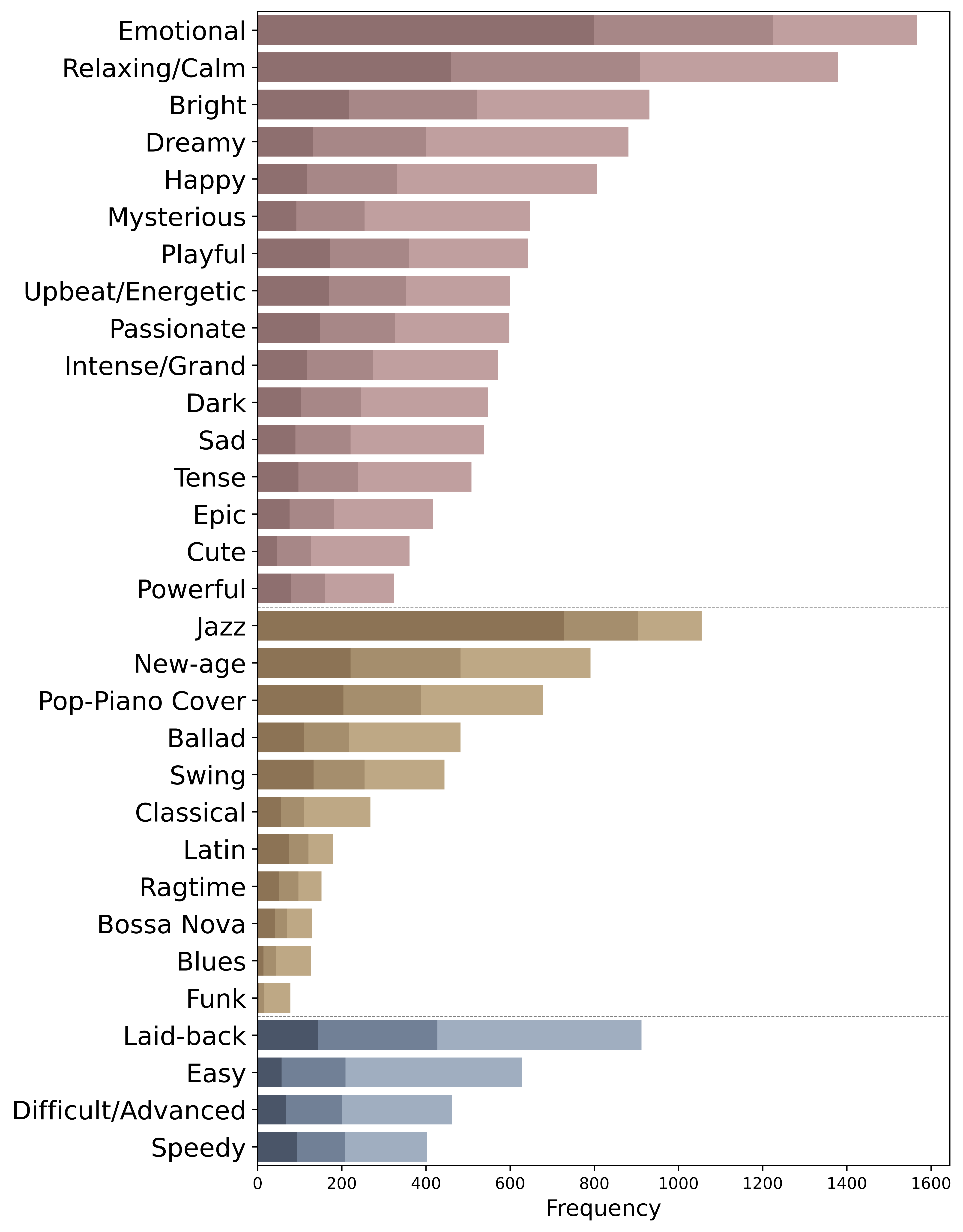}}
 \caption{Tag distribution of the PIAST-AT dataset. Three distinct represent the degree of consensus. (Darkest: n=3, Medium: n=2, Lightest: n=1)
}
\vspace{-2.5mm}
\label{fig:distribution}
 \vspace{-2.5mm}
\end{figure}

Figure \ref{fig:distribution} shows the distribution of tags in the PIAST-AT dataset, categorized into Mood/Emotion, Genre, and Style. The Style category includes tags associated with performance difficulty and tempo-related mood. Due to the inherent imbalance of collected audio, there is also a disparity in the frequency of sub-genre tags. 

The dataset contains the consensus degree among the annotators. To leverage this information, we generated hierarchical captions based on the level of agreement as follows:
\begin{quote}
\vspace{-1mm}
\textit{“This is definitely Jazz genre; (3 agreements)\\
also Speedy style; also Playful mood; (2 agreements)\\
potentially Latin genre; potentially Easy style; potentially Bright, Happy, Cute mood of piano music. (1 agreement)”}
%\vspace{-1mm}
\end{quote}

The PIAST-AT dataset comprises audio, transcribed MIDI, and text annotations (tags and captions), offering a rich representation of musical characteristics and annotator consensus.

% \begin{figure}[t]
%  \centerline{
%  \includegraphics[width=1.0\columnwidth]{fig/annotation_co.png}}
%  \caption{Co-occurrence
%  of the PIAST-AT}
%  \label{fig:distribution}
% \end{figure}

% \begin{figure*}[t]
% \centering
% \includegraphics[width=\textwidth]{fig/tagwise.png}
% \caption{ROC-AUC and PR-AUC scores for each tag in tag-to-music retrieval performance. The darker bars represent audio performance, while the lighter bars represent MIDI performance.}
% \label{fig:tagwise}
% \end{figure*}

\section{Piano Music Classification}

In this section, we present the application of our proposed dataset for piano music annotation and retrieval tasks in both the audio and MIDI domains. We employed a two-stage framework: 1) pre-training and 2) transfer learning. For pre-training, we used the PIAST-YT dataset to train a general-purpose piano-specific model with large-scale audio, MIDI, and diverse text data. We leveraged text supervision through music-text joint embedding pre-training \cite{mulan, muscall, ttmr}. In the transfer learning stage, we utilized the PIAST-AT dataset to train a piano classification model as a downstream task.

\subsection{Pre-training and Transfer Learning}
To develop a piano-specific pre-trained model, we extracted embeddings from audio, MIDI, and text modality encoders. We applied contrastive loss to maximize similarity between corresponding pairs (audio-text or MIDI-text) while minimizing similarity with in-batch negative samples. Following previous studies \cite{mulan, muscall, ttmr}, each encoder consists of a modality-specific backbone, a linear projection layer, and an $l_2$ normalization layer. We used a modified ResNet-50 \cite{resnet} for audio, RoBERTa \cite{roberta} for text, and MidiBERT-Piano \cite{midibert} with average pooling for MIDI.

For the classification model, we employed the probing protocol \cite{toward, codified}. We used the pre-trained audio and MIDI encoders as frozen feature extractors and trained linear models and one-layer MLPs as shallow classifiers on top of them, with 512 hidden states and ReLU activation.

\subsection{Implementation Details}
We processed the input data for pre-training and transfer learning as follows: Audio inputs were 10-second signals sampled at 22050 Hz, converted to log-mel spectrograms with 128 mel bins using a 1024-point FFT with a Hann window and a 10 ms hop size. For MIDI, pre-processed MIDI files were converted to the CP (\textit{compound word}) representation \cite{compound} and fed into a 12-layer BERT with a maximum sequence length of 512. All models were optimized using AdamW with a 5e-5 learning rate, and a dropout rate of 0.4 applied to the audio classification model. We used batch sizes of 128 for audio and 48 for MIDI data during pre-training. Pre-training models were trained for 150 epochs, while classification models ran for 700 epochs, with the best model selected based on validation loss. The PIAST-AT dataset was split into 80\% for training, 10\% for validation, and 10\% for testing sets.

% \textbf{Audio}:
% We followed the implementation details of the original model. The input to the encoder is a 10-second audio signal sampled at 22050 Hz, which is converted to a log-mel spectrogram with 128 mel bins, a 1024-point FFT using a Hann window, and a hop size of 10 ms.\\
% \textbf{MIDI}:
% Following the original MidiBERT-Piano implementation, the transcribed and analyzed MIDI files were converted to the CP representation \cite{compound} and fed into a 12-layer BERT with a maximum sequence length of 512. All hyperparameters were kept the same as in the original model, except for the batch size of the MIDI data.

% Both the data were processed using an RTX A6000 GPU (48GB), with a batch size of 128 for the audio data, and 48 for the MIDI data to fit within the GPU memory. The PIAST-AT dataset was split into 80\% training set, with 10\% used as a validation set and another 10\% used as a test set.

\subsection{Evaluation \& Results}
We evaluated our classification models on two tasks: the \textit{annotation task}, which involves finding appropriate tags for given music, and the \textit{retrieval task}, which focuses on finding suitable music for provided tags. Following previous studies~\cite{zero, musical}, we employed the area under the ROC and PR curves averaged over instances as evaluation metrics for the annotation task. The retrieval task was assessed using the area under the ROC and PR curves averaged over labels. To demonstrate the effectiveness of the proposed large PIAST-YT dataset, we used a supervised model trained exclusively on the smaller PIAST-AT dataset as the baseline model.

\begin{table}[t!]
\centering
\begin{adjustbox}{max width=\columnwidth}
\begin{tabular}{cccccc}
\toprule
& \multicolumn{2}{c}{\textbf{Music $\rightarrow$ Tag}} & \multicolumn{2}{c}{\textbf{Tag $\rightarrow$ Music}} \\
\cmidrule(lr){2-3} \cmidrule(lr){4-5}
 & \textbf{ROC-AUC} & \textbf{PR-AUC} & \textbf{ROC-AUC} & \textbf{PR-AUC} \\
\midrule
\multicolumn{5}{l}{\textit{Supervised}} \\
\textbf{\large Audio} & 81.06 & 70.71 & 73.22 & 50.97 \\
% [1em]
\textbf{\large MIDI}  & 84.82 & 75.24 & 79.14 & 58.00  \\
\midrule
\multicolumn{5}{l}{\textit{Pre-train and Transfer Learning}} \\
\textbf{\large Audio} & 84.52 & 74.73 & 79.01 & 58.70  \\
% [1em]
\textbf{\large MIDI}  & 85.69 & 76.27 & 80.63 & 61.53  \\
\bottomrule
\end{tabular}
\end{adjustbox}
\vspace{-2mm}
\caption{Performance results for music-to-tag and tag-to-music tasks.}
\vspace{-5mm}
\label{table:results}
\end{table}

Table~\ref{table:results} compares the annotation and retrieval performance across 1) audio and MIDI modalities, and 2) the supervised versus pre-train and transfer framework. The MIDI model consistently outperformed the audio model across both tasks. Pre-training with PIAST-YT improved the performance of both models on all metrics, demonstrating its effectiveness. This pre-training approach led to superior performance in both music-to-tag and tag-to-music tasks.

\section{Conclusion}
In this paper, we introduced PIAST, a piano dataset with audio, symbolic and text. Our experiments demonstrated the dataset's effectiveness for piano music annotation and retrieval tasks, showing improvements with pre-training.
The PIAST dataset supports various applications, including improved music retrieval, text-based music generation, music analysis, and emotion/genre classification. To further enhance the dataset, our future work will address tag imbalances by adding more samples and incorporating additional processed data such as lead sheets and chord annotations.

\onecolumn
\section*{Acknowledgments}
This work was supported by the collaboration with NCSOFT, Korea. 

\section*{Appendix}
\vspace{-4mm}
\begin{figure}[hbt!]
    \centering
    \includegraphics[width=0.8\textwidth]{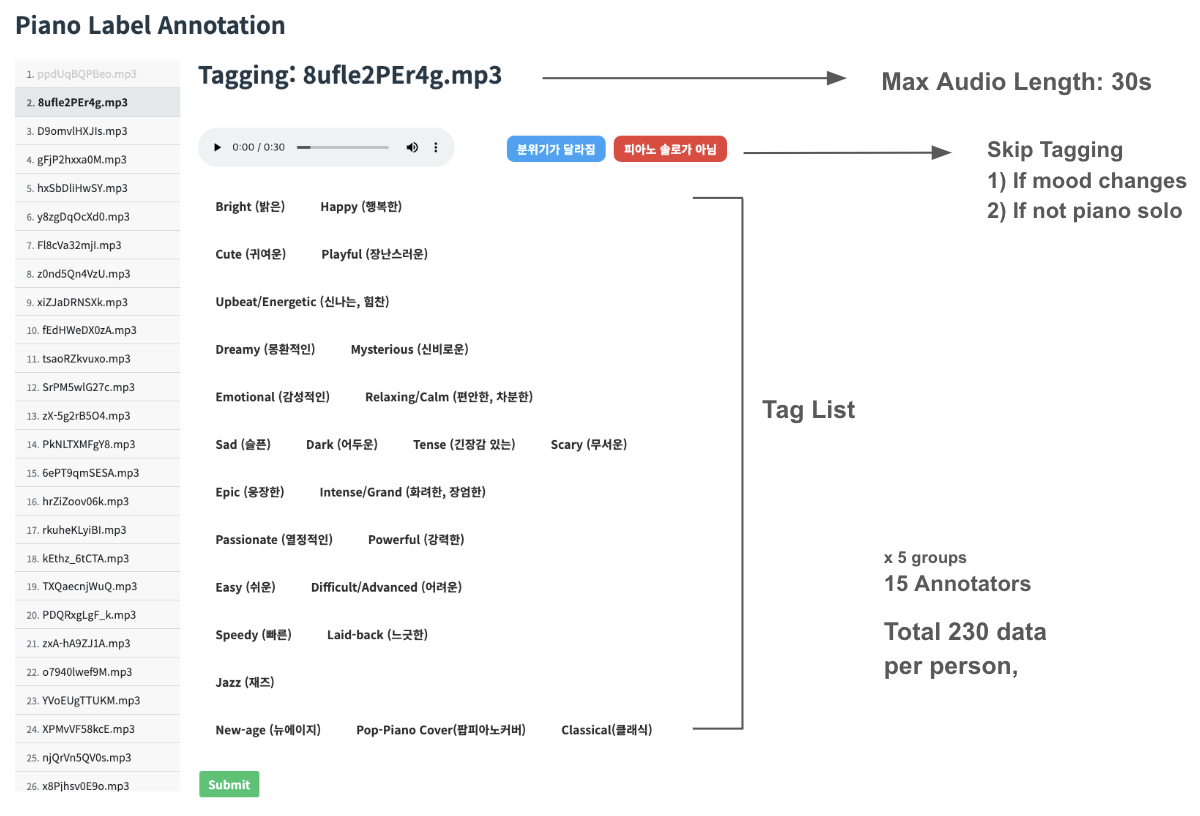}
    \caption{Annotation interface used in the PIAST-AT dataset.}
    \label{fig:annotation}
\end{figure}

\begin{figure}[hbt!]
    \centering
    \includegraphics[width=0.6\textwidth]{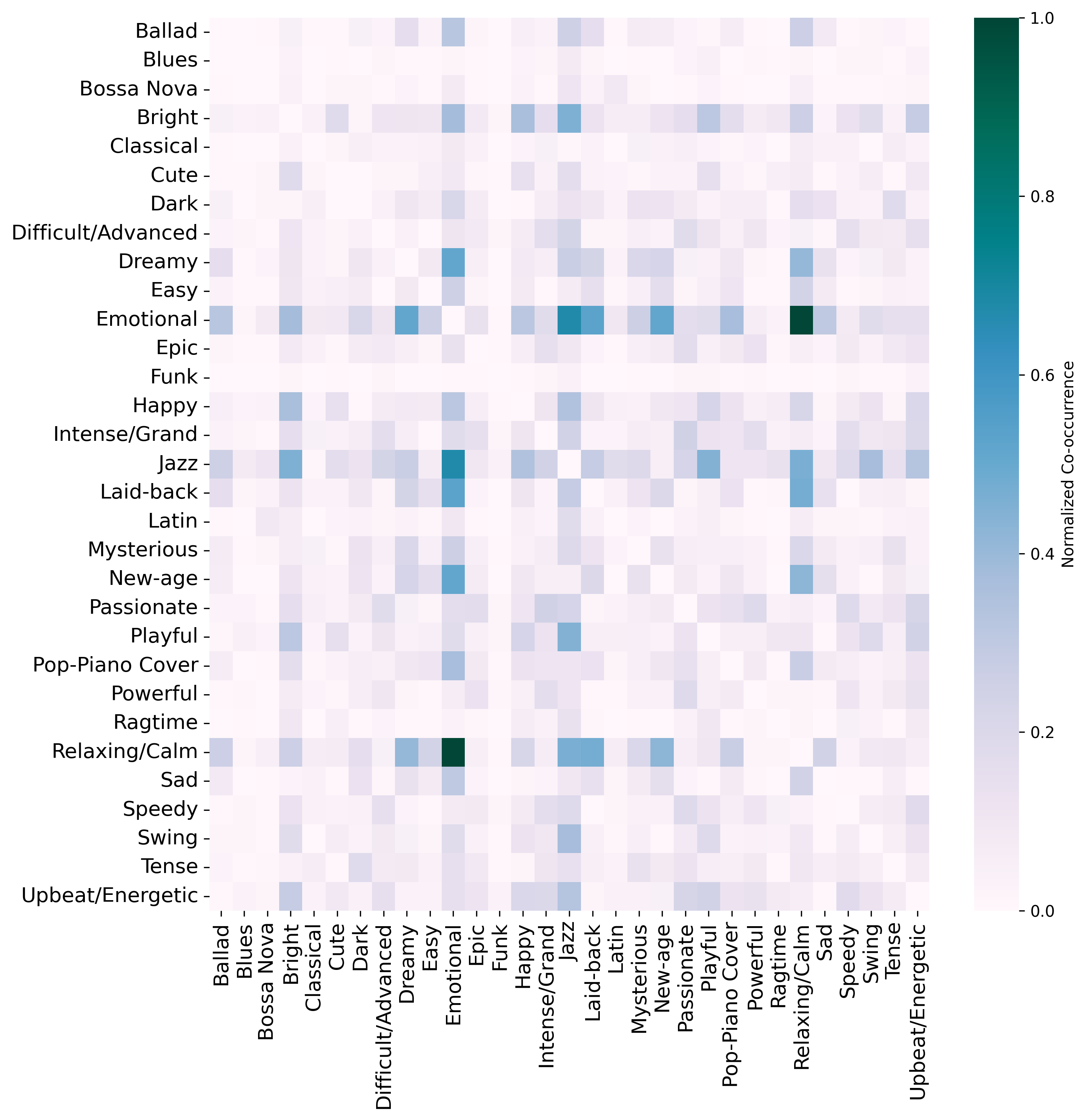}
    \caption{Co-occurrence between tags in the PIAST-AT dataset.}
    \label{fig:co}
\end{figure}
\clearpage

\begin{figure}[hbt!]
    \centering
    \includegraphics[width=1.0\textwidth]{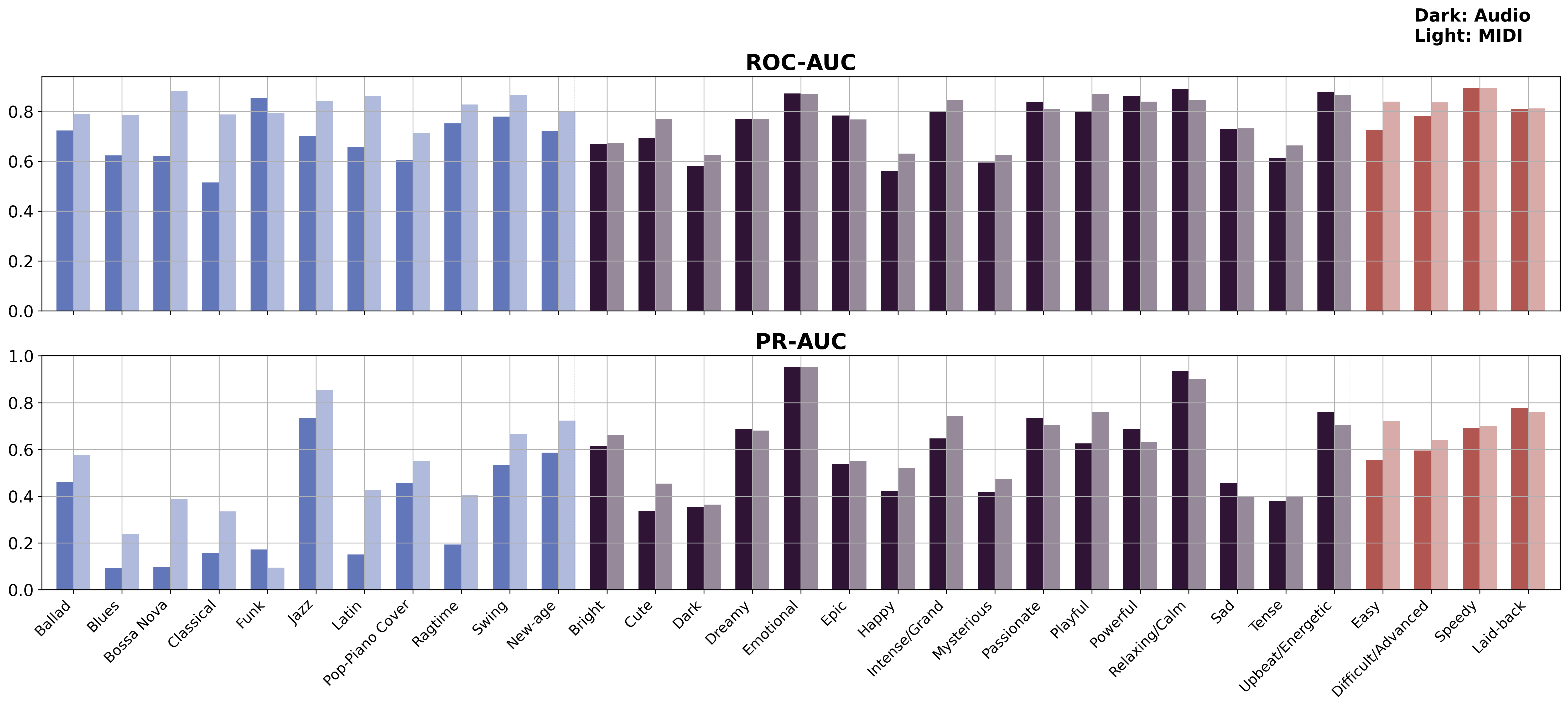}
    \caption{ROC-AUC and PR-AUC scores for each tag in tag-to-music retrieval performance. The darker bars represent audio performance, while the lighter bars represent MIDI performance.}
    \label{fig:tagwise}
\end{figure}

\subsection*{Tag-wise Result Analysis}
 Figure \ref{fig:tagwise} shows the ROC-AUC and PR-AUC scores for both audio and MIDI models across the tags. As shown in Figure \ref{fig:tagwise}, both models exhibited relatively low PR-AUC scores for genre tags. This low performance is likely due to data imbalance, as some genre tags are underrepresented in the PIAST-AT. Despite this imbalance, the MIDI model still performed significantly better than the audio model in most genre tags, suggesting that the MIDI model is more effective in capturing rhythmic characteristics in the music.\\
 For emotion/mood tags and style tags, the performance difference between audio and MIDI was not as pronounced as for genre tags. However, for the ``Cute'' and ``Easy'', the MIDI model is slightly more distinct. This indicates that MIDI data is particularly adept at capturing the nuances associated with those characteristics.

\twocolumn
\bibliography{nlp4MusA}

\end{document}